\documentstyle[prl,amsfonts,twocolumn,aps,epsfig]{revtex}
\begin{document}

\title{From N\'eel   long-range order to  spin-liquids  in  the multiple-spin
exchange model}

\author{W. LiMing$^\dagger$, G. Misguich, P. Sindzingre,  C. Lhuillier}
\address{Laboratoire de Physique Th\'eorique des Liquides-UMR 7600
of CNRS,  Universit\'e  Pierre  et  Marie  Curie,  case 121,   4 place
Jussieu, 75252 Paris Cedex, France\\ E-mail:liming@lptl.jussieu.fr}
\address{$^\dagger$also at Department of Physics, South-China Normal
University, Guangzhou 510631, China\\ (\today)\\}
\maketitle
\bibliographystyle{prsty}
PACS numbers: 75.10.Jm; 75.50.Ee; 75.40.-s
\begin{abstract}
The phase   diagram   of the  multiple-spin  exchange    model on  the
triangular lattice is  studied  using  exact diagonalizations.     The
two-spin ($J_2$) and four-spin  ($J_4$) exchanges have been taken into
account for 12, 16,  19, 21, 24, and  27 site samples in the parameter
region $J_4=0 -  0.25$ (for a  fixed  $J_2=1$).  It is  found that the
three-sublattice N\'eel ordered  state  built up by the  pure two-spin
exchange can    be  destroyed by the  four-spin    exchange, forming a
spin-liquid state.  The different  data suggest that the phase diagram
in this range of parameters exhibits two phases.  The pure $J_2$ phase
is a three-sublattice N\'eel  ordered phase, a  small $J_4$ drives  it
into a spin-liquid  state with a spin gap  filled of a large number of
singlets.  This  spin-liquid phase is not of  the same generic kind as
the phase studied by Misguich  {\it et al.}  [Phys.  Rev.  B {\bf 60},
1064 (1999)].  It is observed on the finite-size samples that the spin
liquid phase, as  the N\'eel ordered  phase,  exhibits a magnetization
plateau at $m = 1/3$,  and for $J_4  >0.15$ a second  plateau at $m  =
1/2$. These  two    plateaus   are associated  respectively    to  the
semi-classical orderings $uud$ and $uuud$.
\end{abstract}
\section{INTRODUCTION}

The two-dimensional triangular  lattice  antiferromagnet (2D-TLA)  was
firstly proposed to be a candidate for the disordered (or spin-liquid)
ground-state of the spin-${\frac 1 2}$ Heisenberg model (Anderson {\it
et al.}  in   70's~\cite{a73,fa74}).  Different approaches  failed  to
support this  conjecture, but favor  a  ground-state with  N\'eel long
range   order  (LRO)~\cite{sgh88,bllp94,cts99}.     Nevertheless,  the
lattice frustration  on  the    2D-TLA  attracts great   interest   of
theorists,   providing  a    challenge for   exotic  antiferromagnets.
Recently,  the   multiple-spin exchange  model   has been  extensively
studied  as  an alternative  to the Heisenberg   model, showing a rich
structure of ground-states~\cite{km97,mblw98,mlbw99}.  In this  model,
the ground-state  can be ferromagnetic (FM),  anti-ferromagnetic (AFM)
with N\'eel  LRO, or a spin-liquid (SL).   A prospective phase diagram
has been  given by Misguich {\it et  al.}, who considered two-, four-,
and five-spin exchange interactions on the 2D-TLA~\cite{mlbw99}.  They
found that a large enough four-spin exchange  interaction drives th FM
phase into a SL phase.  They did not  study how the  AFM N\'eel LRO is
destroyed by    the four-spin   exchange  interaction,  and    how the
transition between N\'eel  LRO and  the  short range RVB  phase  takes
place.  This question is the main object of this paper.

Unhappily there is   no exact method  allowing  the study  of the zero
temperature phases  of such frustrated  systems.  For a finite system,
however, one can  always, in principle,  represent the  eigenstates in
the complete basis of spin configurations.  This allows  one to have a
real touch on  the exact ground-states of  small  size systems through
numerical computations.    The  huge   number  of spin  configurations
($2^N$) becomes a great obstacle on  the way of numerical simulations.
On the most  recent computers, the largest sample  that may be handled
in  exact  diagonalizations has  $6\times6$ sites.   On the triangular
lattice, the Quantum  Monte Carlo method  is plagued by the well-known
sign problem,  but a new  technique  called Stochastic Reconfiguration
allows  handling samples  up  to $12\times12$ sites~\cite{sc99}.   All
these   calculations    point  to  N\'eel    LRO,  with   a sublattice
magnetization of      the  order  of  $40   \%$    of   the  saturated
value~\cite{cts99}.   Series expansions give  a reduced  ($20 \%$) but
non-zero sublattice magnetization~\cite{sh92}.

On  the other  hand, in  the  case  of  short range correlations,  the
situation is more straightforward, as soon  as the available sizes are
of  the  order of, or larger   than  the correlation length.   This is
fortunately the  case in  the SL  phase  found in the  $J_2-J_4$ model
($J_2 \leq 0, J_4 > 0$) by Misguich {\it et al.}~\cite{mlbw99,mblw98}.

In this  work,   we use exact  diagonalizations   to obtain  the exact
eigenenergies versus wave  vectors and total  spin for 12, 16, 19, 21,
24 and  27 site samples of  the $J_2-J_4$ model ($J_2=1,  J_4 > 0$) on
the 2D-TLA. In the classical limit, the AFM ground-state of the 2D-TLA
can  be described   as a three-sublattice    structure, with spins  of
different sublattices  making angles  of $2\pi/3$.  Periodic  boundary
conditions  are compatible   with the  three-sublattice  structure for
samples with 12, 21, 24 and 27  sites, but not for the  16 and 19 site
samples.  Therefore, we use twisted boundary conditions for the 16 and
19 site samples~\cite{bllp94} and periodic boundary conditions for the
12, 21, 24 and 27 site ones.

\section{THE MODEL: THE MULTIPLE-SPIN EXCHANGE HAMILTONIAN}
The Hamiltonian of the multiple-spin exchange model is given by 
\begin{equation}
H=\sum_{n} (-1)^n J_n (P_n+P_n^{-1})\;\;,\; J_n > 0\;\; , n\ge 2
\label{hamilt}
\end{equation}
where   $J_n$  are   the $n$-spin   exchange  tunneling probabilities
(exchange  coefficients),   $P_n$  and  $P_n^{-1}$  are  the  $n$-spin
exchange  operators and their    inverse operators, respectively.  The
alternative  sign in the summation over  $n$ in Eq.~\ref{hamilt} comes
from the  permutation  of   fermions.    In general,   the    exchange
coefficients decrease with increasing  $n$. The two-spin exchange term
gives exactly the  Heisenberg Hamiltonian up to  a constant, since one
has
\begin{equation}
P_2=2 {\bf S}_i \cdot {\bf S}_j + {1\over 2}
\label{exchange}
\end{equation}
where ${\bf s}_i$ and ${\bf s}_j$ are spins localized  at site $i$ and
$j$,  respectively.  The   three-spin exchange   operator  is  exactly
equivalent  to  a  sum of  two-spin  exchange operators~\cite{mlbw99}.
Thus, the three-spin  exchange term of  equation (\ref{hamilt}) can be
absorbed into the two-spin exchange term, as long as $J_2$ is replaced
by    the   effective   two-spin   exchange    coefficient   $J_2^{\rm
eff}=J_2-2J_3$.  Therefore, except for the two-spin exchange, the next
most  important term is   the four-spin  exchange.   A  pure  positive
two-spin exchange   (i.e., the Heisenberg  Hamiltonian)  on the 2D-TLA
gives  an  AFM   phase    with N\'eel     LRO,  and,   as  shown    in
Ref.~\cite{mlbw99}, a  pure four-spin exchange gives  a  SL phase.  In
this paper we  use  the specific properties  of  the spectra of  these
different kinds  of phases to study  the transition from one  phase to
the other,  when the relative  weight of the  four-Spin Exchange $J_4$
increases relatively to  the antiferromagnetic two-spin coupling.   In
the following,  all the energies  are  measured in units  of $J_2^{\rm
eff}=1$.

Little is known on this region of  the phase diagram. Previous works are
based  on    a classical   approximation~\cite{km97},    semi-classical
spin-wave calculations~\cite{ksmn98} or    mean-field  Schwinger-boson
results~\cite{mbl98}.  The classical result predicts a transition from
the  3-sublattice N\'eel state to a  4-sublattice tetrahedral state at
$J_4=0.24$.   Both  quantum  approaches  indicate  that the  four-spin
exchange  strongly enhance   fluctuations in the   3-sublattice N\'eel
phase.  Kubo~{\it et.  al}~\cite{ksmn98}   found that  the  sublattice
magnetization    vanishes  for  $J_4>0.17$.    In the  Schwinger-boson
approach~\cite{mbl98}, the N\'eel state is destroyed when $J_4> 0.25$.
These   two techniques have  a general  tendency  to underestimate the
effects  of  quantum  fluctuations    on ordered phases.    The  exact
diagonalization  analysis presented here  indeed shows that the N\'eel
long-ranged order disappears for a smaller value of $J_4$ (the critical
value is estimated to be in the interval $J_4^C \sim 0.07\cdots 0.1$).

\section{CRITERION TO DISCRIMINATE BETWEEN N\'EEL LRO AND SL
PHASE: THE SPIN GAP~?}
\subsection{Finite-size energy spectrum of the N\'eel LRO phase} 
In  the classical limit,  a $N$-site 2D-TLA sample  with N\'eel LRO is
characterized by a three-sublattice structure  with spin $N/6$ on each
sublattice. Coupling of these  three $N/6$-spins, gives total spin $S$
with  $min\{2S+1,N/2-S+1\}$ degeneracy~\cite{bllp94}.  In an isotropic
antiferromagnet   (as the     collinear  AFM which    has equal    spin
susceptibilities   and spin   wave  velocities) the  finite-size total
energy depends on the total spin $S$ (to first order in $1/N$) as:
\begin{equation}
E_S=E_0+{1\over2 N \chi} S(S+1) ,
\label{rotator}
\end{equation}
where $E_0=N\epsilon_0$ is the energy of the ground-state in
the thermodynamic limit and $\chi$ is the
isotropic magnetic susceptibility of the sample. In the
anisotropic case
this equation should be rewritten:
\begin{equation}
E_S=E_0+{1\over2 N \chi_\perp} S(S+1)+{1\over2 N} ({1\over
\chi_\parallel} - {1\over \chi_\perp}) S_3^2 ,
\label{rotator2}
\end{equation} 
where $S_3$ is the  component of the  total spin  $S$ on the  internal
symmetry  axis  of  the  spin   system,   and $\chi_{\parallel}$   and
$\chi_\perp$  are   the magnetic    susceptibilities on the   internal
symmetry axis and on   the perpendicular plane, respectively.  In  the
broken symmetry picture   the  symmetry axis is  perpendicular  to the
plane of the spins  and $\chi_{\parallel}$ (respectively $\chi_\perp$)
measures   the spin fluctuations orthogonal  to  (respectively in) the
spin  plane.  Eq.~\ref{rotator} (respectively Eq.~\ref{rotator2}),  is
the dynamical  equation of a  rigid rotator (respectively of a quantum
top).

Eqs.~(\ref{rotator}, \ref{rotator2}) show that the slopes of the total
energy versus $S(S+1)$ and  $S_3^2$ approach zero  as $1/N$ does  when
$N\rightarrow\infty$.  $S_3$ is an internal quantum number dynamically
generated, which is not under control in  a finite-size study. But the
total spin $S$ is a good  quantum number and  the $ N^{-1}$ scaling of
the  $S(S+1)$ dependence of  the  total energy  versus sample  size is
interesting because it is more rapid than the scaling law of the order
parameter (which goes as $ N^{-1/2}$)~\cite{bllp94,f89,nz89}.

\subsection{Finite-size scaling in the SL phase}
In a SL phase, contrarily  to the N\'eel  LRO phase, the spin
gap (i.e. the difference in total  energy between ground-states in the
$S=1$ sector and in the $S=0$ sector) does not collapse to zero in the
thermodynamic  limit.  The  finite-size scaling   law in this   second
situation is not known exactly, insofar as  the "massive" phase is not
characterized  precisely.  Heuristically,  we  expect the  finite-size
spin gap to decrease exponentially to a finite value $\Delta (\infty)$
with  the    characteristic      length  $\xi$  of    the    spin-spin
correlations.  For   samples of   linear size   $L$  smaller  than the
correlation length and in the cross-over regime,  there are not enough
quantum fluctuations to  destroy the sublattice magnetization and  the
system probably behaves as  it were classical  (i.e.  with a  spin gap
decreasing as $N^{-1}$).  The following heuristic law might be used to
interpolate between the two behaviors:
\begin{equation}
\Delta (L) =  \Delta (\infty) + \frac {\beta}{L^2} \times exp (- L/\xi) 
\label{gap}
\end{equation}

\subsection{Quantum critical regime}
The use  of  this heuristic law  (Eq.~\ref{gap})  encounters  a severe
difficulty as  soon as  the  disordered  system approaches a   quantum
critical point: in  such  a  situation  the  correlation  length $\xi$
diverges,  the gap closes to  zero and on  a finite-size  sample it is
impossible to   discriminate between such   a  situation and isotropic
N\'eel LRO (Eq.~\ref{rotator}).

\subsection{Numerical results}
\begin{figure}[h]
\centerline{\psfig{figure=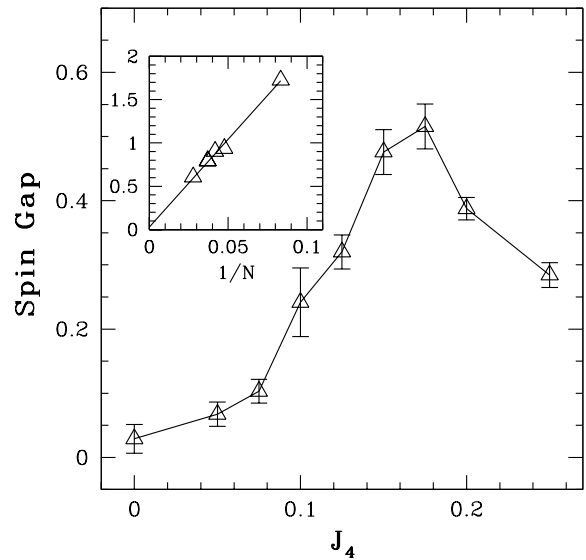,width=8.4cm}} 
\caption[99]{Extrapolation $N\rightarrow\infty$ of the spin gaps of 12, 21, 24
and  27  site  samples for   different  $J_4$.  The  inset  shows  the
extrapolation at $J_4=0$.   The spin gap for  $N=36$ is evaluated from
data of Ref.~\cite{bllp94}.    Since not  all symmetry sectors    were
investigated for $S=1$ and $N=36$, the energy of the lowest triplet is
not   known  exactly.   Therefore  we     estimated the  spin  gap  by
$\Delta\simeq\frac{1}{6}     \left[E(S=3)-E(S=0)\right]$,      as   in
Ref.~\cite{cts99}.  Lines are guides for the eye.}
\label{spingap}
\end{figure}

In view  of this  difficulty we  have   done a pedestrian  finite-size
scaling of  the spin gap by using  the  simplest linear $1/N$ behavior
which probably gives a lower bound of the gap in a SL outside critical
points.   The  physical   reason  is that   we  expect the finite-size
corrections of the gap  value to be smaller  in a system with a finite
correlation  length (no long-range order) than  in a  LRO N\'eel phase
where it vanishes as $1/N$.~\footnote{This assumption would be invalid
at a critical point where the uniform susceptibility vanishes: In such
a   case the  gap  might  close   as  $1/\sqrt{N}$.  The  analysis  of
section~\ref{sec:sym} shows  that the behavior  of the  system changes
rather abruptly from a N\'eel like spectrum  to a spectrum with a very
large  number of low  lying singlets below  the first $S=1$ state (and
potentially  a  $T=0$ residual entropy  in  the singlet sector).  This
happens  before any decrease to zero  of the spin  velocity  or of the
homogeneous spin  susceptibility.    So, if the  true  thermodynamical
system has indeed  a critical point between the  two phases, the sizes
we  are looking at are  too small to  scrutinize the critical regime.}
The   results  extrapolated to    $N\rightarrow  \infty$ are shown  in
Fig.~\ref{spingap}.  Strictly speaking the spin gap never extrapolates
to zero except for a pure  $J_2$ where it  is equal to zero within its
error bar (At  $J_4=0$, data of  Bernu~{\it  et al.}~\cite{bllp94} for
$N=36$ are added   to the present results, see   Fig.~\ref{spingap}.).
Nevertheless these  data  already show three  distinct  ranges for the
parameter $J_4$:  for  very small $J_4$   (below 0.075) N\'eel  LRO is
plausible but   should be confirmed by  another  approach.   For $J_4$
larger than $0.1$ a gap  certainly opens rapidly with increasing $J_4$
and then decrease  for $J_4 >  0.175$.  The spin gap criterion  cannot
give more   insight on the  phase diagram.   We will  now move  to the
analysis of the symmetries  of the low  lying levels of the spectra to
characterize more precisely these three phases.

\section{SYMMETRIES OF THE LOW LYING LEVELS IN A N\'EEL ORDERED
PHASE}\label{sec:sym}
\subsection{Theoretical background}
\begin{figure}
\begin{center}

\centerline{\psfig{figure=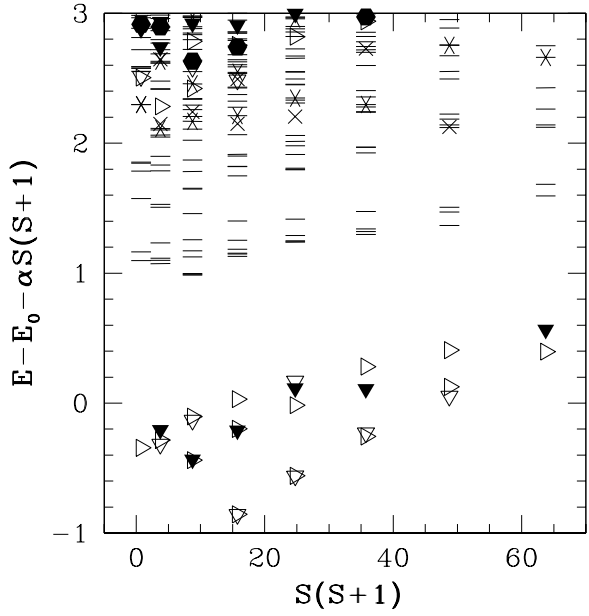,width=8.4cm}} 

\caption{Spectrum of the Heisenberg model ($J_4=0$)for the 21-site sample.
$\bigtriangledown$: levels   with    ${\bf  k}=(0,0)$ and   symmetries
$R_{2\pi/3}=1$ and $R_\pi=1$ ($R_\theta$  is the phase factor obtained
in  a      $\theta$-rotation   about   the    origin);     black solid
$\bigtriangledown$: levels    with   ${\bf k}=(0,0)$   and  symmetries
$R_{2\pi/3}=1, R_\pi=-1$; $\rhd$: levels with ${\bf k}_0$ (the corners
of the Brillouin zone) and symmetries $R_{2\pi/3}=1$; $-$: levels with
wave vectors inside the Brillouin zone.}
\label{spect21}
\end{center}

\end{figure}

Firstly we  show the low energy spectrum  of the pure Heisenberg model
on the 21 site sample (Fig.~\ref{spect21}).  In order to emphasize the
low energy structure we have displayed the low energy spectrum minus a
rigid rotator energy $ \alpha S(S+1)$.

Let us first concentrate on the lowest part  of the energy spectrum in
each S sector (solid and open triangular symbols in the figure).  This
family of levels     forms  on a  finite-size  lattice     the quantum
counterpart of the semi-classical N\'eel state.  These specific states
are in the   trivial representation  of  the invariance  group  of the
three-sublattice  N\'eel ordered     solution~\cite{bllp94}.   To   be
definite:
\begin{itemize}
\item The N\'eel ground-state breaks the 1-step translation but is
invariant in a  3-step translation:  as a  consequence  the only  wave
vectors      appearing    in      this   family     of      QDJS  (for
Quasi-degenerate-joint-states   defined   by   Bernu    {\it        et
al.}~\cite{bllp94}) are respectively  the center $  {\bf k}=(0,0)$ and
corners $
\pm {\bf k_0}$ of the Brillouin zone. 
\item These QDJS belong specifically to the trivial representation
of $C_{3v}$ as the  N\'eel state itself (i.e.  they are invariant in a
$2\pi /3$ rotation, and in a reflection symmetry).
\item As the $\pi$ rotation symmetry of the lattice is broken in
this particular ground-state, these  QDJS appear either  in the odd or
even  representation      of   the  2-fold    rotation     group  (see
Ref.~\cite{bllp94} for more details).
\item The numbers and characteristics (quantum numbers) of the QDJS in
each $S$ sector are  precisely fixed by theory~\cite{bllp94}: for  the
21  sites spectrum displayed in  Fig.~\ref{spect21}  (as for all sizes
that have been studied up to now) the numbers of  low lying levels and
their quantum   numbers  correspond  exactly to  the   above-mentioned
theoretical predictions.
\end{itemize}
\subsection{Numerical results}
The dynamical law  given  by Eq.~\ref{rotator2} is still   imperfectly
obeyed for  the 21 site  sample: in  particular the  generation of the
internal symmetry is still imperfect  but nevertheless the spectrum of
a quantum top could already be anticipated.

Above these levels with specific  properties, there appear eigenstates
with wave   vectors  belonging to  the inside  of   the Brillouin zone
(simple dashes in  Fig.~\ref{spect21}).  A  group of such  eigenstates
with different total spin represents a magnon excitation of the N\'eel
ground-state.   As    the  antiferromagnetic magnons     have a linear
dispersion law, the softest magnon energy  scales as the smallest wave
vector  accommodated in the  Brillouin zone of the finite-size sample.
Thus,   for  sizes large     enough, these   levels  collapse   to the
ground-state as $1/\sqrt N$, more slowly  than the QDJS which collapse as
$1/N$  to the thermodynamic N\'eel  ground-state  energy.  That is the
reason of the appearance in  Fig.~\ref{spect21} of a quasi gap between
the QDJS and the magnon excitations.

This hierarchy of low lying levels is a very  strong constraint on the
finite-size samples spectra. It is perfect for sizes up  to 27 and for
$J_4$  smaller  or equal  to   0.075 and  totally  absent~\footnote{An
illustration of  the striking qualitative differences  between spectra
in the N\'eel region and in the spin-liquid  phase has been previously
published: two $N=27$ spectra  at $J_4=0$ and $J_4=0.1$  are displayed
in Fig.~2 and Fig.~3  of  Ref.~\cite{mlbw99}.}   for $J_4$ larger   or
equal to 0.1.

\begin{figure}
\begin{center}

\centerline{\psfig{figure=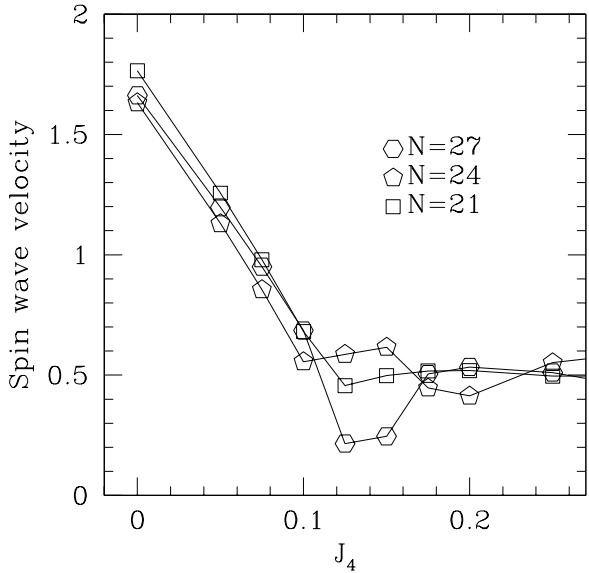,width=8.4cm}} 

\caption{Spin wave velocities of 21, 24 and 27 -sites samples. Lines are
guides for the eye. These quantities are computed as  the ratio of the
first $\Delta S=1$ excitation  energy divided by  the momentum of  the
corresponding excitation  (according  to $v=\Delta E/\Delta{\bf  k}$).
For $J_4>0.1$,  these   numbers  do not correspond   to   well defined
physical excitations  because  of  the  strong perturbations  of   the
spectra.  Notice  at  $J_4=0.1$, Fig.~\ref{singlets}  exhibits a  huge
number of low-energy singlet states below  the spin gap and the system
is therefore already in a SL phase.}
\label{spinwave} 
\end{center}
\end{figure}

\begin{figure}
\begin{center}

\centerline{\psfig{figure=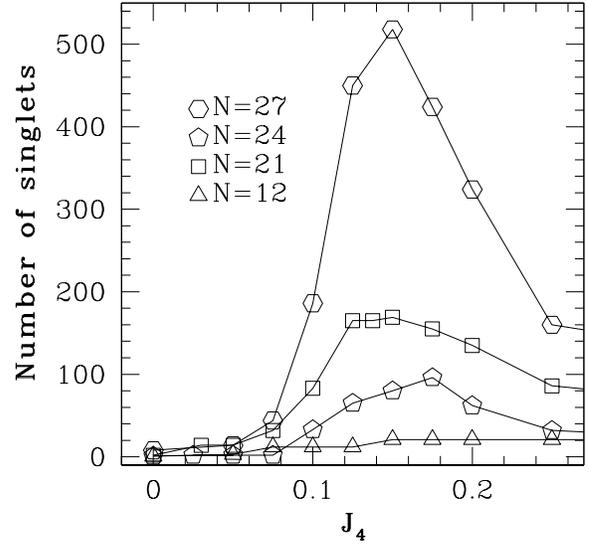,width=8.4cm}} 

\caption{Number of singlet states between the ground-states of the $S=0$
(or  $1/2$) and $S=1$ (or  $3/2$)  sectors.  Lines are  guides for the
eye.}
\label{singlets}
\end{center}
\end{figure}

\begin{itemize}
\item This  result, 
associated to  the  spin gap  behavior, consistently  proves  that for
$J_4$ larger or equal to 0.1 the system is in a spin-liquid state with
rather short range spin-spin correlations.
\item For $J_4 \le 0.075$  the structure of the low lying
eigen-levels of the  spectra are compatible with  N\'eel LRO.  BUT  as
discussed above, it is indeed impossible to  precisely point a quantum
critical transition within this approach.  In view  of the spectra, we
might speculate that the transition is second order  and that it takes
place between 0.07 and 0.1.
\item
From  $J_4=0$ to $J_4\simeq0.1$ we   see a softening of the  spin-wave
velocity consistent  with  the  gradual  decrease  of the N\'eel   LRO
(Fig.~\ref{spinwave}).  However,    Fig.~\ref{singlets} shows that   a
large number of singlet states are  already present at low energy when
$J_4=0.1$.  Therefore,  the  system  is certainly   no longer  in  the
ordered    phase  at     $J_4\simeq0.1$.    We   can   conclude   from
Fig.~\ref{spinwave} that the spin wave velocity does not vanish at the
critical point (even if the precise  location of the transition cannot
be   determined).   This  has  been  previously   suggested by various
analytical approaches~\cite{adm92,css94,sss94}.  The sizes studied are
nevertheless too small to check Azaria's prediction~\cite{adm92} of an
$O(4)$ symmetry of the effective field theory at the critical point.
\end{itemize}

\section{THE LOW ENERGY EXCITATIONS OF THE SL PHASE}

Contrarily to our expectations, the SL phase which appears immediately
after the disappearance  of the N\'eel ordered phase  is not the phase
studied by Misguich~{\it et  al.}~\cite{mlbw99}.   It is indeed  a  SL
phase, with a  gap and  short range  spin-spin correlations.  But,  as
might be seen in Fig.~\ref{singlets}, this phase exhibits a very large
number of  singlets in the   magnetic gap and   seems in this  respect
similar to the  spin-liquid   phase of the   Heisenberg model  on  the
kagom\'e  lattice~\cite{lblps97,web98}.  However, we  are not aware of
any   exponential degeneracy  ({\it  i.e.}   $\sim  \exp(N)$)  in  the
classical MSE model~\footnote{Momoi~{ \it et al.}~\cite{msk99} found a
ground-state degeneracy in the classical MSE model but this degeneracy
only grows   as    the exponential  of linear    size  of  the  system
($\sim\exp{\sqrt{N}}$).   Moreover, this classical degeneracy was only
found    in a  region    where    $J_2<0$   is {\em     ferromagnetic}
($\frac{1}{4}\leq  J_4/|J_2|\leq \frac{3}{4}$).},  as  is the case for
the classical kagom\'e antiferromagnet.

We suspect that a  much larger  four-spin  exchange parameter  will be
needed to recover the SL phase studied by Misguich {\it et al.}. These
data point to  the existence of  this new phase in  a finite  range of
parameters $0.075   \leq J_4\leq0.25$.  However, one  cannot disregard
the hypothesis that these properties  are in fact  those of a critical
point, with a  critical region enlarged by  finite-size effects.  More
work with different methods is needed to clarify this point.
 
\section{Magnetization Plateaus}

In an external magnetic field $B$ along the $z$ axis, the total energy
of the state with component $S_z$ of the total spin is given by:
\begin{equation}
E_B=E_S - S_z B
\label{EB}
\end{equation}
The magnetization is determined by the minimum of $E_B$ respective to
$S_z$, 
which requires $\partial E_B/\partial S_z=0$. Therefore, in the isotropic
case with N\'eel LRO, one has
\begin{equation}
m=2\chi B-1/N
\label{magnetization}
\end{equation}
where  $m=2S/N$, is  the  polarization   relative  to  the   saturated
magnetization $N/2$ and $\chi$ is indeed the magnetic susceptibility.
   
\begin{figure}
\begin{center}

\centerline{\psfig{figure=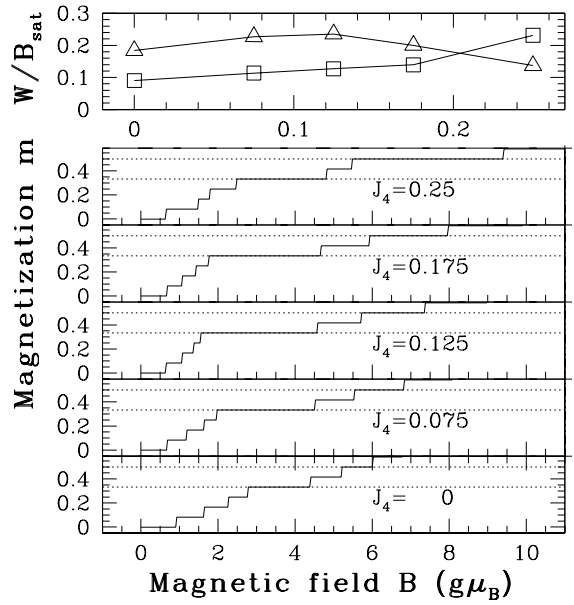,width=8.4cm}} 

\caption{Magnetization versus external magnetic field for different  
$J_4$ ($N=24$).  The    dotted   lines  label  the $1/3$     and $1/2$
magnetization.   The  top  panel   shows   the  widths  of   the $1/3$
($\bigtriangledown$) and $1/2$ ($\Box$) plateaus versus $J_4$}
\label{plateau}
\end{center}
\end{figure}

We noticed  that  when the  four-spin exchange interaction  increases,
deviation from Eq.~\ref{rotator}   occurs about $S_z=N/6$  and  $N/4$.
This is in agreement with  the earlier mean-field calculation of  Kubo
and Momoi~\cite{km97} who predicted  magnetization plateaus at $m=1/3$
and $m=1/2$   in the $J_2-J_4$   model.  We present  the magnetization
curve  of the 24-site sample in  Fig.~\ref{plateau}.   A small plateau
exists  at  $1/3$  magnetization  for $J_4=0$,   and  its  width first
increases slowly as  $J_4$  increases and  then  decreases from around
$J_4=0.125$.    This $1/3$ plateau   has  also been found in  previous
studies  (see~\cite{h99}    and references     therein) in  the   pure
three-sublattice  N\'eel  ordered  system.  The $m=1/2$  magnetization
appears  at  about $J_4=0.1$  and  its  width  increases   with $J_4$.
Finally, there still exists a plateau at about $m=1/2$ in samples with
odd  numbers   of  sites, but it   distributes   over the two  closest
positions to  the    $1/2$ magnetization.   The   $m=1/3$  and $m=1/2$
plateaus correspond  to the  classical $uud$  and  $uuud$  ordering of
spins (see Momoi  {\it  et al.}~\cite{msk99} for   the four-sublattice
$uuud$ state).  These phases are  more "classical" than the zero-field
phase, but yet show a   decrease of the sublattice magnetization  from
the classical saturation values.

\section{Conclusion}
In  conclusion, we studied the  transition between  the N\'eel ordered
state and  a Spin Liquid state of  the multiple-spin exchange model by
means of the  exact diagonalization method.  The pure three-sublattice
N\'eel ordered phase is   gradually destroyed by  quantum fluctuations
when increasing the 4-spin exchange coupling.   The spin wave velocity
decreases but  apparently   remains finite  at  the  transition.   The
quantum disordered phase tuned  by   the 4-spin exchange  coupling  is
different from the pure $J_4$  phase studied by Misguich~{\it et al.}.
It exhibits low energy singlet  excitations, reminding of kagom\'e SL.
This result opens many  interesting questions that cannot  be answered
in the  present framework: is this  phase a new generic  SL phase or a
finite-size manifestation of a quantum   critical regime~?  Are  these
singlet excitations the   ``resonon'' modes  invoked by Rokhsar    and
Kivelson~\cite{rk88}~?          In   agreement   with         previous
studies~\cite{km97}, we find two magnetization  plateaus at $1/3$  and
$1/2$ of the full magnetization.  These plateaus are associated to the
semi-classical $uud$   and $uuud$ ordering  structures.  A finite-size
scaling on much larger  sizes is needed to  draw a definite conclusion
on this magnetic phase diagram.

\newpage
\end{document}